\documentclass[aps,showpacs,onecolumn]{revtex4}
\usepackage{epsfig}

\begin{document}

\title{Problems related to gauge invariance and momentum, spin decomposition in nucleon structure study
\footnote{Corresponding author: fgwang@chenwang.nju.edu.cn (Fan Wang)}}

\author{FAN Wang$^{a}$, W.M. SUN$^{a}$, X.S. Chen$^{b}$ and P.M. Zhang$^{c}$}

\affiliation{$^a$Department of Physics and CPNPC (joint Center for Particle Nuclear Physics and Cosmology),
  Nanjing University and Purple Mountain Observatory, Chinese Academy of Science, Nanjing 210093, China}

\affiliation{$^b$Department of Physics, Huazhong Science and Technology University, Wuhan 430074, China}

\affiliation{$^c$Institute of Modern Physics, Chinese Academy of Science, Lanzhou 730000, China}

\begin{abstract}
How do the quark and gluon share the nucleon momentum?
How does the nucleon spin distribute among its constituents?
What means the quark and gluon momentum, spin and orbital angular momentum?
These problems are analyzed and a solution is proposed based on {\it gauge
invariance principle, canonical quantization rule and Poincar$\acute{e}$ covariance}.
\end{abstract}

\pacs{}

\maketitle

\setcounter{totalnumber}{6}

\section{Introduction}

Nucleon is the dominant component of visible mass in the universe. It is a
fundamental laboratory for the study of the microscopic structure of matter controlled
by strong interaction and the low energy scale properties of strong interaction
theory, the nonperturbative QCD. The quark model proposed in 1964 suggested
a naive picture of the nucleon internal structure. It is assumed to
consist of three valence quarks, all of them occupying the lowest s-wave orbit.
Therefore the nucleon spin is solely due to quark spin. This picture explains
the anomalous nucleon magnetic moments, discovered in 1933, quite well.
The lepton-nucleon deep inelastic scattering (DIS) in 1960's-1970's confirmed that
there are really colored, spin one-half, fractional charged particles within the nucleon.
However, the polarized $\mu-p$ deep inelastic scattering (DIS) measurement in
1987 showed that the quark spin contribution is only a small amount of nucleon spin
and this led to the so-called "proton spin crisis". After the experimental
effort of a quater of century, it is confirmed that the quark spin contribution is
about $0.25(\frac{\hbar}{2})$. How does the nucleon spin distributes
among its constituents is still a controversial issue.
Even how do the quark and gluon share the nucleon momentum is also under
debate. We found that there have been confusions about what are the quark and gluon
momentum, spin and orbital angular momentum which complicated the
nucleon structure study. In this report we will analyze these problems
and propose a solution based on {\it gauge invariance principle, canonical
quantization rule and $Poincar\acute{e}$ covariance}.

\section{The first proton spin crisis and the quark spin confusion}

M. Gell-Mann and G. Zweig proposed the quark model to describe the nucleon internal structure
in 1964. In this model the nucleon is assumed to consist of three valence quarks
which are all located in the lowest s-wave orbital state. The spin-flavor part is assumed to be
in an $SU^{f\sigma}(6)\supset SU^{f}(3)\times SU^{\sigma}(2)$ symmetric state. Based on this
picture, the nucleon spin is solely due to quark spin contribution. The quark orbital
angular momentum (OAM) does not contribute. The gluon spin and OAM do not contribute, either.
For proton, the quark spin contributions are,
\begin{eqnarray}
\Delta{u}=\frac{4}{3},\Delta{d}=-\frac{1}{3},\Delta{s}=0, \nonumber \\
L_q=L_G=\Delta{G}=0,
\end{eqnarray}
where $\Delta{u}, \Delta{d}, \Delta{s}, \Delta{G}$ stands for flavor u, d, s quark
and gluon polarizations, $L_q, L_G$ are the quark and gluon OAM contributions.
In 1988, the European Muon Collaboration (EMC) published their polarized
$\mu-p$ DIS result, which shows a surprisingly small quark spin contribution~\cite{EMC},
\begin{equation}
  \Delta{\Sigma}=\Delta{u}+\Delta{d}+\Delta{s}\sim{0}.
\end{equation}

There had been a long time debate on what this result means~\cite{AELC}. Experimentalists
spent 25 years to improve the measurement continuously. The present overall fit result is~\cite{FL},
\begin{eqnarray}
\Delta{\Sigma}=0.817-0.453-0.055=0.255(Q^2=1GeV^2,\Delta{G}=-0.118), \nonumber\\
\Delta{\Sigma}=0.814-0.456-0.056=0.245(Q^2=4GeV^2,\Delta{G}=-0.096), \nonumber\\
\Delta{\Sigma}=0.813-0.458-0.057=0.242(Q^2=10GeV^2, \Delta{G}=-0.084),
\end{eqnarray}
where the numbers in the above equations are $\Delta{u},\Delta{d},\Delta{s}$ respectively
and $\Delta{G}$ is gluon spin polarization. Theoretically, the gluon anomaly contribution
to the matrix element (ME) of the flavor singlet axial charge can be absorbed into the quark
spin contribution in the gauge invariant factorization scheme. Therefore the popular idea is
that the "proton spin crisis" is standing and the quark model picture of nucleon structure is
invalidated by the polarized lepton-nucleon DIS measurements.

We did an analysis of the nucleon spin puzzle in 1998~\cite{DCW}, where we pointed out:
(1) the polarized DIS measured "quark spin" contribution is the ME of the
flavor singlet quark axial vector current operator for a longitudinal polarized proton.
Even though this axial vector current operator is the relativistic field-theoretical extension
of the non-relativistic Pauli spin operator, it is different from the latter which is
calculated in quark model to obtain the quark spin contribution. One should calculate
the ME of the axial vector current operator which includes the relativistic
correction and $q\bar{q}$ creation (annihilation) terms in addition to the Pauli spin.
(2) the pure valence quark $q^3$ configuration is only the dominant component even for a
ground state proton, and there is sea quark excitation or the QCD vacuum polarization component
in the ground state proton. One should take these sea-quark components into account in a
refined quark model calculation.

Based on this we did a Fock space expansion quark model calculation
of the ME of the axial vector current operator and found that this model predicts the
quark spin content correctly with almost the same model parameters as used in the standard
ones. Moreover, this model reproduces all of the masses and magnetic moments of the ground state
octet and decuplet baryons. In this model, the relativistic correction shifts $\sim{0.32}$
quark spin to quark OAM. The ${q^3}\rightleftharpoons {q^3}q\bar{q}$ crossing
ME due to the $q\bar{q}$ creation(annihilation) shifts $\sim{0.44}$ quark spin
to quark OAM. The u,d,s quark magnetic moment modifications almost
cancel each other and leave the non-relativistic quark model predicting the correct magnetic
moments of the ground state baryons. It also gives the correct ${g_A}=\Delta{u}-\Delta{d}=1.23$.
The numerical results of the u,d,s quark polarizations are term by term consistent with
the lattice results~\cite{Liu}. The physical mechanism is also the same, our model and lattice
calculation both show that it is the disconnected and connected diagrams which shift quark spin
to OAM. We had done a transformation to transform our model u,d,s quark momentum
distributions in the rest frame to the infinite momentum frame by the method developed in~\cite{DMZ}
and found that qualitatively the model parton distributions are similar to the measured ones for
low $Q^2$, especially in the valence quark kinematical region.

Then where does the nucleon get its spin? Our model answer is that the relativistic
field-theoretical quark OAM will compensate the quark spin losing. The relativistic quark spin
$\bf{S}_q$ (the axial vector current operator) and OAM operator $\bf{L}_q$ can be expanded as
follows through the decomposition of the Dirac spinor into its large and small components,
\begin{eqnarray}
\bf{S}_q&=&{\int}d^3x{\psi}^{\dag}\frac{\bf{\Sigma}}{2}{\psi} \nonumber \\
&=&\frac{1}{2}{\int}d^3x\bar{\psi}{\bf {\gamma} }{\gamma}^5{\psi} \nonumber\\
&&={\sum}_{i,\lambda,{\lambda}'}{\int}
d^3k{\chi}^{\dag}_{\lambda}\frac{\bf{\sigma}}{2}{\chi}_{{\lambda}'}
(a^{\dag}_{i,\bf{k},\lambda}a_{i,\bf{k},{\lambda}'}
-b^{\dag}_{i,\bf{k},{\lambda}'}b_{i,\bf{k},\lambda}) \nonumber \\
&&-\frac{1}{2}{\sum}_{i,\lambda,{\lambda}'}{\int}d^3k
{\chi}^{\dag}_{\lambda}
\frac{\bf{\sigma}{\cdot}\bf{k}}{k_0(k_0+m_i)}i\bf{\sigma}{\times}
\bf{k}{\chi}_{{\lambda}'}
(a^{\dag}_{i,\bf{k},\lambda}a_{i,\bf{k},{\lambda}'}
-b^{\dag}_{i,\bf{k},{\lambda}'}b_{i,\bf{k},\lambda}) \nonumber \\
&&+{\sum}_{i,\lambda,{\lambda}'}{\int}d^3k{\chi}^{\dag}_{\lambda}
\frac{i\bf{\sigma}{\times}\bf{k}}{2k_0}
{\chi}_{{\lambda}'}a^{\dag}_{i,\bf{k},\lambda}b^{\dag}_{i,-\bf{k},
{\lambda}'}+H.C..
\end{eqnarray}
\begin{eqnarray}
\bf{L}_q&=&{\int}d^3x{\psi}^{\dag}\bf{x}{\times}
\frac{\nabla}{i}{\psi} \nonumber \\
&&={\sum}_{i,\lambda}{\int}d^3k
(a^{\dag}_{i,\bf{k},\lambda}i\bf{\nabla}_k{\times}\bf{k}a_{i,\bf{k},
\lambda}+b^{\dag}_{i,\bf{k},\lambda}i\bf{\nabla}_k{\times}
\bf{k}b_{i,\bf{k},\lambda}) \nonumber \\
&&+\frac{1}{2}{\sum}_{i,\lambda,{\lambda}'}{\int}d^3k
{\chi}^{\dag}_{\lambda}
\frac{\bf{\sigma}{\cdot}\bf{k}}{k_0(k_0+m_i)}i\bf{\sigma}{\times}
\bf{k}{\chi}_{{\lambda}'}
(a^{\dag}_{i,\bf{k},\lambda}a_{i,\bf{k},
{\lambda}'}-b^{\dag}_{i,\bf{k},{\lambda}'}b_{i,\bf{k},\lambda})
\nonumber \\
&&-{\sum}_{i,\lambda,{\lambda}'}{\int}d^3k
{\chi}^{\dag}_{\lambda}\frac{i\bf{\sigma}{\times}\bf{k}}{2k_0}
{\chi}_{{\lambda}'}a^{\dag}_{i,\bf{k},\lambda}b^{\dag}_{i,-\bf{k},
{\lambda}'}+H.C..
\end{eqnarray}
Eq.(4) shows what we already mentioned above that the relativistic quark spin
includes three terms, the third line is the usual non-relativistic Pauli spin
which is what one calculated in the usual non-relativistic quark model to obtain
the quark spin contribution expressed in Eq.(1), the fourth line is the relativistic
correction which shifts quark spin to quark OAM, the fifth line is the $q\bar{q}$
creation(annihilation) term which will couple the dominant valence $q^3$ component
to the minor valence quark core and meson cloud components $q^3q\bar{q}$ and shift
further the quark spin to the OAM. The relativistic reduction of quark spin
had been calculated in many relativistic quark models. However, the
$q\bar{q}$ creation(annihilation) term contribution has almost never been
calculated in the vast quark model calculations even after the "proton spin crisis"
and this is the most important mechanism (the disconnected and connected diagrams)
to shift the quark spin to OAM~\cite{AT}.

Comparing Eq.(4) and (5), one can find that the relativistic correction and $q\bar{q}$
creation(annihilation) terms in the relativistic field theoretical-quark spin and OAM
cancel each other exactly and so we have,
\begin{equation}
\bf {S}^{nr}_q+\bf {L}^{nr}_q=\bf {S}^r_q+\bf {L}^r_q.
\end{equation}
This means that the relativistic quark OAM contribution will compensate the quark spin
losing exactly. If we assume the static gluon field does not contribute to nucleon spin
and the quantum fluctuation of the background gluon field is small, then the nucleon spin
can be viewed as mainly due to the non-relativistic quark spin $\bf {S}^{nr}_q$ and
a very small amount of non-relativistic quark OAM $\bf {L}^{nr}_q$ because of the small
sea-quark component or a small amount of the relativistic quark spin $\bf {S}^r_q$ and a
large amount of the relativistic quark OAM $\bf {L}^r_q$ contributions. Eq.(4),(5) and (6)
also shows that the non-relativistic spin and OAM operators used for a long time in the
non-relativistic quantum mechanics, if added together, is correct. Of course,
the relativistic field-theoretical ones are more realistic.

The non-linear realization of chiral symmetry spontaneous breaking
\begin{eqnarray}
\tilde{\psi}(x)&=&e^{i{\gamma}^5{\xi}_a(x){\lambda}_a}\psi(x)
\end{eqnarray}
transforms the current quark to the constituent quark and this
transformation keeps the quark vector and axial vector current
operators invariant~\cite{MG}. This serves as the non-perturbative QCD basis of
our model numerical calculation.

Based on these analysis we conclude that there is no "proton spin crisis" and the quark model picture
of the nucleon structure is qualitatively correct. But there is "quark spin confusion" in
misidentifying the relativistic field-theoretical spin to the non-relativistic Pauli spin!

\section{The second proton spin crisis and the quark orbital angular momentum confusion}

R.L. Jaffe talked about the second proton spin crisis in the {\it 1st International Symposium
on Science at J-PARK}, Mito, Ibaraki, Japan, 2008, where he compared the quark model
OAM with the lattice calculated and the "measured" quark OAM through the generalized
parton distribution (GPD). A.W. Thomas made a quantitative comparison of their quark
model OAM contribution and the lattice calculated and GPD "measured" one~\cite{AT}. Such
kind of comparisons are popular in the literature. However, these are in fact quite different
quantities. The quark model calculated one is the ME of the non-relativistic
{\it canonical} quark OAM $\bf {L}^{nr}_q$ or the relativistic {\it canonical} one $\bf {L}^r_q$.
The existing lattice calculated one or the "measured" one through GPD is the ME of the
following operator,
\begin{eqnarray}
\bf {L}_{qk}&=&\int d^3x\psi^{\dag}\bf {x}\times\frac{\bf{D}}{i}\psi, \nonumber \\
\bf{D}&=&\nabla-ig\bf{A}.
\end{eqnarray}
This "OAM" $\bf{L}_{qk}$ includes the gauge potential $\bf{A}(x)$ of
the studied physical system. For example, within the nucleon
or meson the $\bf{A}(x)$ is different and so the $\bf{L}_{qk}$ for nucleon
and meson will have different meaning. Moreover, this operator does not satisfy the
angular momentum algebra,
\begin{equation}
\bf{L}_{qk}\times\bf{L}_{qk}\neq i\bf{L}_{qk}.
\end{equation}
So the present lattice calculated and the measured one through GPD are not the
canonical OAM and to compare them with what had been calculated in quark models
is nonsense. Even such an "OAM" is measured completely many years later, it is
useless for checking our qualitative picture about the nucleon internal spin distribution
obtained from quark models. If we misidentify this "measured quark OAM" to the
quark model calculated one, it will lead to the second "proton spin crisis" as
already happened in the literature.

\section{what is the gluon spin?}

It is widely believed that one can measure the gluon spin contribution to nucleon spin
through the measurement of the gluon parton helicity distribution. It is also widely
believed that there is no gauge invariant local operator corresponding to the first moment
of gluon helicity parton distribution. Even more generally, there is no gauge invariant
local operator corresponding to the spin of a massless particle including gluon and photon.
On the other hand, the optical community believed that they had measured the photon spin.
In all of the microscopic structure study, from atom to hadron, the multipole radiation
analysis is crucial where the spin and OAM of photon is unavoidable. All of these call
for a spin operator for the massless particles, the photon and gluon.

\section{Decomposition of the total momentum and angular momentum of a gauge field system into its constituents}

In the study of the nucleon structure, it is essential to understand how the mass,
the momentum, the angular momentum or the spin distribute among its constituents. The
above discussions show that one has to do a critical analysis on what are the momentum,
spin and orbital angular momentum of quark and gluon first.
Atom is a QED system and nucleon is a QCD system, they are both gauge field systems
and therefore the above problems are common to atom and nucleon. We will choose the
simple QED system to analyze the above problems and propose a solution based on {\it gauge
invariance principle, canonical quantization rule and $Poincar\acute{e}$
covariance} first. The results are then extended to QCD case.

R.L. Jaffe and A. Manohar first obtain a decomposition of the total angular momentum
of the QCD gauge system into quark and gluon spin and OAM parts in 1990~\cite{JM},
\begin{equation}
\bf{J} = \int d^3 x \psi ^\dagger \frac 12 \bf{\Sigma} \psi + \int
d^3x \psi ^\dagger \bf{x} \times\frac 1i \bf{\nabla} \psi + \int d^3x \bf{E}\times \bf{A}
+\int d^3x E^i\bf{x}\times \bf{\nabla} A^i.  \label{JM}
\end{equation}
This expression can be applied both to QED and QCD, while for QCD case, the color
indices are omitted. The first term is the electron (quark) spin, the second term
is the electron (quark) OAM, the third term is the photon (gluon) spin, and the
fourth term is the photon (gluon) OAM. The advantage of this decomposition is
that each term satisfies the canonical angular momentum algebra and so is qualified to
be called electron (quark) spin, OAM and photon (gluon) spin, OAM. The disadvantage
of this decomposition is that only the first term is gauge invariant and all the other
three terms are gauge dependent and so not measurable.

To remedy this drawback, our group and X.D. Ji obtained another decomposition in
1997~\cite{CWJI}
\begin{equation}
\bf{J} = \int d^3 x \psi ^\dagger \frac 12 \bf{\Sigma} \psi + \int
d^3x \psi ^\dagger \bf{x} \times\frac 1i \bf{D}\psi +\int
d^3x \bf{x}\times \left(\bf{E}\times\bf{B}\right)
\end{equation}
This expression can be applied both to QED and QCD and the color indices are omitted
again. The advantage of this decomposition is that each term is gauge invariant. The
disadvantage is that the individual term does not satisfy the canonical angular momentum
algebra except the first electron (quark) spin term. So to call them electron (quark)
OAM $\bf{L}_{qk}$ and total angular momentum of photon (gluon) at least will cause confusion.
This really happened as we discussed in the third section. Moreover, such kind of OAM is a
mixing of electron and electro-magnetic potential in QED case and a mixing of quark and gluon potential
in QCD case. For different systems, different atom or molecule in QED, different
baryon or meson in QCD, such an OAM will include different gauge potential contribution.
The $\bf{L}^2_{qk}, \bf{L}_{z,qk}$ do not commute and so cannot be diagonalized
simultaneously. Another disadvantage is that the total
photon (gluon) "angular momentum", i.e., the third term, has not been further decomposed into
photon (gluon) spin and OAM. For a long time it is widely believed that it is impossible to
further decompose the total angular momentum of a massless particle into gauge invariant
spin and OAM. This conflicts with the photon spin measurements in optics and the on going
gluon spin measurement. It is also inconsistent with the widely used multipole radiation
analysis where both photon spin and OAM are inevitable. Despite these serious drawbacks
this decomposition has been widely accepted as a benchmark for the study of
nucleon spin structure. However, we never suppose
it is a good decomposition for the angular momentum of a gauge field system!

Instead we had turned to another direction to develop the A.S. Wightman's idea: what quantum
measurement measures is the ME of an operator. Gauge dependent operator might have gauge
invariant ME for physical states~\cite{SW}. If this can be proved, then one can use the
Jaffe-Manohar gauge dependent decomposition to study the nucleon spin structure.
In order to go beyond the Lorenz gauge, we use path-integral formalism to study this
possibility~\cite{CWJI}, because Wightman's argument is limited to the Lorenz gauge.
Unfortunately in the course of this study we found that the path-integral formalism does
not always give reliable result, mainly because it involves divergent integrals.~\cite{SCW}
Therefore, whether a gauge dependent operator could have gauge invariant ME for physical
states is still an open question.
Then we turn back to search for a decomposition, in which each term is individually
{\it gauge invariant}, satisfies {\it canonical quantization rule} and satisfies
{\it Poincar$\acute{e}$ covariance} as much as possible. This is possible both for QED and QCD.
Let us discuss the simpler QED case first. The Abelian $U(1)$ gauge potential in QED case
can be decomposed as follows,
\begin{equation}
A_{\mu}(x)=A_{\mu,phs}(x)+A_{\mu,pure}(x),~~ A_{\mu,phys}(x)=\frac{1}{{\partial}^2} \partial_i F_{i\mu}(x),
~~A_{\mu,pure}(x)=A_{\mu}(x)-A_{\mu,phys}(x).
\end{equation}
Under suitable boundary condition the solution of Eq.(12) is unique. It is easy to check,
\begin{equation}
F_{\mu \nu} = \partial_{\mu} A_{\nu,phys}-\partial_{\nu} A_{\mu,phys}, ~~
F_{\mu \nu,pure}= \partial_{\mu} A_{\nu,pure}-\partial_{\nu} A_{\mu,pure}=0,
\end{equation}
\begin{equation}
\partial_i A_{i,phys}=\partial_i\frac{1}{{\partial}^2}\partial_k F_{ki}=0,~~
\partial_i A_{i,pure}=\partial_i A_i.
\end{equation}
The spatial part of $A_{\mu,phys}$ and $A_{\mu,pure}$ are nothing else but the
transverse and longitudinal part of the well known Helmholtz decomposition.
Under a local gauge transformation,
\begin{eqnarray}
\psi'(x)&=& exp(-ie\omega(x))\psi(x), \nonumber \\
A'_{\mu}(x)&=&A_{\mu}(x)+\partial_{\mu}\omega(x),
\end{eqnarray}
The $A_{\mu,phys}$ and $A_{\mu,pure}$ will transform as follows,
\begin{equation}
A'_{\mu,phys}(x)=A_{\mu,phys},~~~A'_{\mu,pure}=A_{\mu,pure}+\partial_{\mu}\omega(x).
\end{equation}
Please note the $A_{\mu,phys}$ is gauge invariant!
Based on these properties of $A_{\mu,phys}$ and $A_{\mu,pure}$, we obtain the following
decomposition,
\begin{eqnarray}
\bf{J}&=&\int d^3 x \psi ^\dagger \frac 12 \bf{\Sigma} \psi + \int
d^3x \psi ^\dagger \bf{x} \times\frac 1i \bf{D}_{pure} \psi \nonumber \\
&+& \int d^3x \bf {E} \times \bf {A}_{phys}+ \int d^3x E^i\bf{x}\times \bf{\nabla} A_{phys}^i.
\end{eqnarray}
Here $\bf {D}_{pure}$ is,
\begin{equation}
\bf{D}_{pure}=\bf{\nabla}-ie\bf{A}_{pure}.
\end{equation}
The $\frac{\bf{D}_{pure}}{i}$ is the gauge invariant extension of the canonical
momentum which reduces to the canonical momentum in the $\bf{A}_{pure}=0$ gauge,
the Coulomb gauge. The three components of $\frac{\bf{D}_{pure}}{i}$
commute with each other, the same as the canonical momentum. The commutator between
this operator and the OAM, the second term in the above equation, is the same as those
of canonical momentum and OAM in the Poincar$\acute{e}$ algebra. Due to these properties
we call them the "physical momentum". In fact we can obtain a corresponding decomposition
of the total momentum of the QED system,
\begin{equation}
\bf{P}=\int d^3x\psi^{\dag}\frac{\bf{D}_{pure}}{i}\psi+\int d^3x E^i\bf{\nabla}A^i_{phys}.
\end{equation}
It is not hard to check that the individual term in these two decompositions satisfies
gauge invariance and the canonical momentum and angular momentum quantization rule.
This is our proposed decomposition of the momentum and angular momentum for the QED system.
It is the first one to decompose the
total angular momentum of a massless particle into a gauge invariant spin (the third term)
and OAM (the fourth term) operator and provides
the gauge-field theoretical basis for the optical measurements of photon spin and OAM
and the widely used multipole radiation analysis from atomic to hadron spectroscopy.
In general this decomposition includes non-local operator if one uses Eq.(12) to
calculate the $A_{\mu,phys}$. However in Coulomb gauge, where $A_{\mu,pure}=0$ and
$A_{\mu,phys}=A_{\mu}$, one no longer needs to use Eq.(12) to calculate the $A_{\mu,phys}$,
this decomposition will include local operator only. In fact in Coulomb gauge our
decomposition reduces to Jaffe-Manohar decomposition Eq.(10). Therefore one can use the simple
Jaffe-Manohar definition of electron (quark) spin, OAM and photon (gluon) spin, OAM to
do calculations in Coulomb gauge. In this case, all of these operators have their
familiar canonical forms used in quantum mechanics. This also explains why we get
the correct results in the vast atomic, molecular, optical, nuclear, hadronic
(including our Fock space extension quark model) calculations even though superficially
gauge dependent operators have been used. It is also a warning: to use these gauge
dependent operators to do calculations beyond Coulomb gauge is dangerous! The time
dependent gauge transformation dependence of the eigenvalue calculation of the hydrogen
Hamiltonian is a typical example~\cite{SG}.

This decomposition can be extended to QCD, the $SU(3)$ non-Abelian gauge field
system. To simplify the expressions we will omit the color index and Gell-Mann
$SU(3)$ color matrix $\frac{\lambda^a}2$ except a few necessary cases.
The gauge potential $A_{\mu}(x)=A^a_{\mu}(x)\frac{\lambda^a}2$,
is decomposed in the same way as in Eq.(12),
\begin{equation}
A_{\mu}(x)=A_{\mu,phys}(x)+A_{\mu,pure}(x).
\end{equation}
The physical condition for the separation of the $SU(3)$ non-Abelian gauge potential
is different from the Abelian $U(1)$ case due to the complication of non-linearity,
\begin{equation}
F_{\mu\nu,pure}=\partial_{\mu}A_{\nu,pure}-\partial_{\nu}A_{\mu,pure}+ig[A_{\mu,pure},A_{\nu,pure}]=0,
\end{equation}
\begin{equation}
D^{adj}_i A_{i,phys}=\partial_i A_{i,phys}+ig[A_{i,pure},A_{i,phys}]=0,
\end{equation}
\begin{equation}
\partial_i A_{0,phys}=\partial_i A_0-\partial_0 A_{i,pure}+ig[A_{i,pure},A_{0,pure}],
\end{equation}
here the braket $[..,..]$ is the commutator of the color $SU(3)$ matrix. We cannot get
the concise solution as in the QED case but we have found perturbative solution~\cite{CWW}.
Under local gauge transformation,
\begin{eqnarray}
\psi'(x) &=& U(x)\psi(x), \nonumber \\
U(x) &=& exp(-ig\omega^a(x)\frac{\lambda^a}2), \nonumber \\
A'_{\mu}(x) &=& U(x)A_{\mu}(x)U^{\dag}(x)-\frac{i}{g}U(x)\partial_{\mu}U^{\dag}(x),
\end{eqnarray}
the $A_{\mu,phys}$ and $A_{\mu,pure}$ transform as follows,
\begin{eqnarray}
A'_{\mu,phys}(x) &=& U(x)A_{\mu,phys}(x)U^{\dag}(x), \nonumber \\
A'_{\mu,pure}(x) &=& U(x)A_{\mu,pure}U^{\dag}(x)-\frac{i}{g} U(x)\partial_{\mu}U^{\dag}(x).
\end{eqnarray}
Based on these properties of $A_{\mu,phys}$ and $A_{\mu,pure}$ we obtain a decomposition
of the total angular momentum $\bf{J}$ of the QCD system. The expression is the same as
Eq.(17) if the color indices are omitted but the last term, where the $\nabla$ should be
replaced by the $\bf{D}^{adj}$ in Eq.(22).
\begin{eqnarray}
\bf{J}&=&\int d^3x\psi^{\dag}\frac{\bf{\Sigma}}{2}\psi+\int
d^3x \psi^{\dag}\bf{x}\times\frac{\bf{D}_{pure}}{i}\psi
\nonumber\\
&&
 +\int d^3x \bf{E}\times\bf{A}_{phys}+\int d^3x
E^i\bf{x}\times\bf{D}^{adj}A^i_{phys}.
\end{eqnarray}
The total momentum of QCD system can be decomposed as follows,
\begin{equation}
\bf{P}=\int d^3x\psi^{\dag}\frac{\bf{D}_{pure}}{i}\psi+\int d^3x E^i\bf{D}^{adj}A^i_{phys}.
\end{equation}
This momentum decomposition is consistent with the angular momentum decomposition given
in the above equation. In these momentum and angular momentum decompositions, each term is
gauge invariant and satisfies part of the Poincar$\acute{e}$ algebra for the three momentum
and OAM components.

Let us digress a little bit but still a related problem.
There is also a long standing problem in non-relativistic and relativistic
quantum mechanics that the fundamental operators, the momentum, OAM and Hamiltonian of a
charged particle moving in electro-magnetic field do not have gauge invariant ME. We
proposed to use the following operators,
\begin{equation}
p^{\mu}-eA^{\mu}_{pure},~~\bf{x}\times(\bf{p}-e\bf{A}_{pure}),
\end{equation}
as the four momentum operator and OAM. This proposal does not change the Dirac equation
for the electron or quark and the Maxwell equation for the photon or gluon field. However,
the interpretation of the Dirac equation is modified a little bit,
\begin{equation}
(i\gamma^{\mu}((\partial_{\mu}+ieA_{\mu,pure})+ieA_{\mu,phys})-m)\psi=0,
\end{equation}
we call $\frac{\partial_{\mu}+ieA_{\mu,pure}}{i}$ as four "physical momentum", the interaction
term only includes the physical part of gauge potential $A_{\mu,phys}$. The advantage of this
proposal is that the four "physical momentums" now all have gauge invariant ME and so are measurable.
One can choose the $A_{\mu,pure}$ to transform as the same as the $\partial_{\mu}$ under a
Lorentz transformation, then the "physical momentum" will transform as a normal four vector. The
three spatial components of momentum and OAM still satisfy the canonical commutation relations.
The disadvantage is that they are no longer the space-time translation generator of the electron or quark
field. Only in a special gauge, the $A_{\mu,pure}=0$ Coulomb gauge, they reappear as the translation
generator. It is impossible to have full Poincar$\acute{e}$ algebra for the electron (or quark) operators
among themselves, especially the three momentum do not commute with the Hamiltonian.
On the other hand, if we take the $P^{\mu}-eA^{\mu}$ as four momentum (they are called
kinetic momentum in the literature), as in the other decomposition depicted in Eq.(11),
\begin{equation}
\bf{P}=\int d^3x \psi^{\dag}\frac{\bf{D}}{i}\psi+\int d^3x \bf{E}\times\bf{B}.
\end{equation}
Even though each term is gauge invariant, but no one satisfies the canonical momentum algebra.
They are no longer the space translation generators, do not satisfy the full Poincar$\acute{e}$
algebra, either. Especially, because the three components
of the quark kinetic momentum do not commute, they cannot be diagonalized simultaneously
and so cannot form a complete momentum operator set to describe the three dimensional quark
parton momentum distribution. The Dirac equation appears to be a
free particle equation and indeed it appears as a free particle momentum in the light-cone gauge.
This has caused confusions to identify this kinetic momentum as canonical momentum. The other
weak point is the Poynting vector $\bf{E}\times\bf{B}$, even it had been identified as the momentum density of photon for a long time, it is in fact not the right momentum density because the $\bf{x}\times(\bf{E}\times\bf{B})$ is
not the OAM but the total angular momentum of photon which includes both photon spin and OAM.

Eq.(17),(19) and (26),(27) are our proposed solution of the momentum and angular momentum
decomposition~\cite{CLSW}. If we keep the gauge invariance requirement, the Jaffe-Manohar
decomposition depicted in Eq.(10) is excluded. If we require to keep the canonical quantization
rule further, the Chen-Wang and Ji decomposition depicted in Eq.(11) and (30) will be excluded,
too. A gauge invariant spin operator of massless particle is obtained in our proposed
decomposition. M. Wakamatsu proved the total angular momentum of gluon (photon) in the Chen-Wang
and Ji decomposition can be decomposed further into spin and OAM and the obtained gauge invariant
spin operator is the same as the one given in Eq.(17) and (26)~\cite{WA}. So the remaining problem is which
momentum and OAM should be chosen? Different choices will give different physical picture of the quark
gluon contribution to nucleon (also different electron photon contribution to atom) observables~\cite{CSLW}.

From atomic structure to hadron spectroscopy one always use the momentum and angular momentum expressed in
Eq.(10) and their gauge invariant version Eq.(17) and (20). The popular idea, however, is that
the kinetic momentum and angular momentum expressed in Eq.(11) and (30) are the
right choice to describe the quark, gluon momentum and angular momentum measured in DIS. They are derived
from the symmetric, gauge invariant Belinfante energy-momentum tensor. It is also widely believed that the
density of the energy momentum tensor (DEMT) is fixed only up to a surface term and one has the freedom to
choose either the canonical asymmetric or Belinfante symmetric one. How large freedom do we really have? The DEMT
of the classical scalar field of the hydrodynamics is measurable and it is symmetric. The DEMT of the classical
electro-magnetic field and the Dirac electron field should be measurable too. In the Belinfante version,
$T^{\mu\nu}_{el}$ (electron part) and $T^{\mu\nu}_{ph}$ (photon) part both are gauge invariant
and in principle measurable. In the usual canonical DEMT  both the electron and photon part are
gauge dependent and so not measurable. However we have derived a new version,
\begin{eqnarray}
T^{\mu\nu}&=& T^{\mu\nu}_{el}+T^{\mu\nu}_{ph}, \nonumber \\
T^{\mu\nu}_{el}&=&\frac{i}{2}\bar{\psi}\gamma^{\mu}(\partial^{\nu}+ieA^{\nu}_{pure})\psi+h.c. \nonumber \\
T^{\mu\nu}_{ph}&=&-F^{\mu\rho}\partial^{\nu}A_{\rho,phys}-g^{\mu\nu}\mathcal{L}_{ph}.
\end{eqnarray}
This is the $T^{\mu\nu}$ which we used to derive our momentum and angular momentum decomposition Eq.(17)
and (19). Both electron and photon parts are gauge invariant and in principle measurable. We therefore
suggest to measure the DEMT of the classical electron and photon field which might be helpful in making the
choice discussed above.

Let us discuss an ideal experiment to illustrate this point. Suppose we have a spin polarized
and orbital unpolarized electron
beam propagating along the z direction. To do a proper discussion one should use a wave packet. But to simplify
the discussion we still use plane wave approximation. We measure the total $J_z$ carried by this beam in a
fixed volume. It should be the total number N of electron in this volume multiplied by the spin $S_z$ of
electron,
\begin{equation}
\int dV J_z(x)=NS_z.
\end{equation}
If we start from the Belinfante symmetric DEMT the above equation can be reexpressed as,
\begin{equation}
\int dV J_z(x)=\int dV (\bf{x}\times{\bf{p}}(x)_{el})_z.
\end{equation}
Due to the symmetric property of the Belinfante DEMT, the momentum density in the above equation is equal
to the energy flow density. The energy flow density divided by the energy $\varepsilon$ of individual electron,
which should be the same for a plane wave, should equal the electron number density flow $n(x)$. This
electron number density flow should equal the electric current density divided by the charge e of electron.
Taking these relations into account we can reexpress the above equation further as,
\begin{equation}
\int dV ({\bf x}\times{\bf p}(x)_{el})_z = \frac{\varepsilon}{e}\int dV ({\bf x}\times{\bf j})_z.
\end{equation}
The last integral should equal the electron number N in this volume multiplied by the spin magnetic moment
$\mu_z$ of electron but with a factor 2, which equals $2\frac{e}{\varepsilon}S_z$ and finally we have,
\begin{equation}
\int dV (\bf{x}\times\bf{p}(x)_{el})_z=2NS_z.
\end{equation}
This obviously contradicts the result shown in Eq.(32) and it originates from the symmetry property
of the Belinfante DEMT,
\begin{equation}
T^{0i}_{el,B}=T^{i0}_{el,B}=\frac{{T^{0i}_{el,C}+T^{i0}_{el,C}}}{2}.
\end{equation}
Here the subscript B and C mean Belinfante and canonical DEMT. The Belinfante momentum (energy flow)
density is a mixing of the physical canonical momentum and energy flow density. In the canonical
version, the $\bf{x}\times\bf{p}$ is related to the OAM density only and will not cause such a
contradiction because the gyromagnetic ratio is 1 for orbital magnetic moment to OAM not 2 as in the
spin case. The fundamental spin can never be related to the orbital motion. The Belinfante DEMT
expresses the spin and OAM superficially both as $\bf{x}\times\bf{p}$ which is physically misleading.

Such an unphysical feature also appears in photon case. In the Belinfante version, the total angular
momentum of photon is expressed as $\bf{x}\times(\bf{E}\times\bf{B})$. Our optical colleagues already showed
experimentally that the spin and OAM of photon are different. D.P. Ghai, S. Senthikumaran and R.S.
Sirohi showed a diffraction pattern of orbital polarized light beam with $L_z=\pm 1$. The pattern is distorted due
to the transverse orbital motion~\cite{GSS}. Our group measured a diffraction pattern of spin polarized
light beam with $S_z=\pm 1$. There is no distortion because there is no orbital motion related to photon spin~\cite{CXS}. However, in the Belinfante description, they are both $J_z=\pm 1$ light beam!

All of these results show that the Belinfante symmetric DEMT might be not physical and the canonical
asymmetric one might be physical. The surface term manipulation is unavoidable in decomposing the
momentum and angular momentum into contributions of constituents for a gauge field system. The
question is to what extent such a manipulation still gives physical result. We propose to measure
the DEMT of gauge field system to test its symmetry property. One example is to use the existed
longitudinal spin polarized electron beam to realize the above ideal experiment.


\section{Discussions}
Since our proposal of the decomposition of momentum and angular momentum of gauge field systems
there have been very intense studies and hot debates on this topic. E. Leader and C. Lorc$\acute{e}$
(L.L.) gave an in time critical review in arXiv: 1309.4235[hep-ph]. Many problems have been clarified
there. We feel the following problems need to be studied further:

(1)L.L. give a new definition of gauge invariance and gauge independence which are different from
what had been given by Wightman~\cite{SW}. We suggest it is better not to do so if it is not
absolutely necessary because it will cause confusion in an already full of confusions field. Up
to now we can not prove rigorously that a gauge dependent operator might have gauge invariant ME
for some physical states, we'd better keep the requirement that an operator corresponding to
an observable should be gauge invariant under the assumption that the physical states are gauge
invariant as in the QCD case.

(2)For QED there are only two photon helicity states observed in Compton scattering and
in Coulomb gauge only these two components of the electro-magnetic potential are kept.
In the other popular Lorenz or light-cone gauge the un-physical components remain.
In this sense Coulomb gauge is special but certainly this situation does not contradict with
the spirit of gauge invariance. On the contrary, gauge invariance will not change the fact that
there are only two physical components for the gauge potential. The Aharonov-Bohm effect is
also induced by this transverse components of electro-magnetic potential non-locally produced
by the electric current within the local solenoid~\cite{LS}. The well known gauge principle
introduces pure and physical gauge potential together but one can have a gauge invariant formalism
without physical interaction,
\begin{equation}
(i\gamma^{\mu}(\partial_{\mu}+ieA_{\mu,pure})-m)\psi=0.
\end{equation}
Here the pure gauge potential is introduced solely for the Dirac equation to be gauge invariant
under a local phase change of the Dirac field. This pure gauge potential can be eliminated by
a local phase change of the Dirac field. On the other hand the transverse components, the
physical part of the gauge potential is not changed under such a local gauge transformation
and can not be eliminated by any gauge transformation.
Our condition to separate the gauge potential into physical and pure gauge parts is based on
this physics. The Helmholtz separation just meet this requirement. There are infinite
possibilities to separate the gauge potential into gauge invariant and variant parts but
no one is able to separate the physical and un-physical parts so clear-cut. In this sense
the physical and pure gauge separation is unique. For QCD case we extend the transverse
condition to meet the gauge covariance also taking into account the fact that there should
be only two helicity components for gluon. But our physical condition is reduced to
transverse condition only in $A^{\mu}_{pure}=0$ case and the gauge covariance transformation
of $A^{\mu}_{phys}$ will mix the un-physical part into physical one in addition
to Gribov ambiguity. So it should be studied further~\cite{ZP}

(3)L.L. already explain the general Lorentz covariance in their review paper. Because there is
mis-understanding of the Lorentz covariance in the literature we suppose it is worth to discuss
this problem here a little bit. The Lorentz transformation of four coordinates, four momentum,
electro-magnetic field tensor are determined by measurements and it is the well known
Lorentz transformation law (LTL). Hereafter we call it homogeneous LTL.
However the transformation law of the gauge potential is not
measurable because of the gauge degree of freedom. J.D. Bjorken and S.D. Drell, and S. Weinberg
already used the more general LTL for the Coulomb gauge potential~\cite{BDW}.
Hereafter we call it in-homogeneous LTL. The popular mis-understanding is to
require the gauge potential to follow the homogeneous LTL no matter which gauge is,
otherwise one supposes that it violates the Lorentz covariance or says it is not manifestly
Lorentz covariant. A typical example is in X. Ji's comment, he criticized that the physical
condition $\nabla\cdot\bf{A}(x)_{phys}=0$ violates the Lorentz covariance~\cite{J}. In fact it is
impossible to prove that gauge potential should transform with homogeneous LTL in general.
Even for the gauge potential in Lorentz gauge it is still possible to transform in-homogeneously
if the residual gauge degree of freedom is taken into account. Of course usually one prefers
to assume it transform with homogeneous LTL. For $\bf{A}_{phys}$, if one assumes it
to transform with homogeneous LTL, the un-physical components will mix in and so one needs
to do an additional gauge transformation to eliminate the un-physical components,
\begin{equation}
A'^{\mu}_{phys}(x')=\Lambda^{\mu}_{\nu}(A^{\nu}_{phys}(x)+\partial^{\nu}\Omega(\Lambda)).
\end{equation}
This new $A'^{\mu}_{phys}(x')$ will be still physical, i.e.,
\begin{equation}
\nabla'\cdot\bf{A}'_{phs}(x')=0.
\end{equation}
The in-homogeneous term $\partial^{\nu}\Omega(\Lambda)$ is not hard to fix and certainly
one doesn't need to do Lorentz boosting! In general the gauge potential transform with
in-homogeneous LTL.

(4)We suppose what means "measurable" might need further study. First we have to distinguish
classical measurable and quantum measurable. X. Ji and M. Wakamatsu argued that the kinetic
or mechanical momentum $m\bf{v}$ is measurable. We think it is classically measurable but quantum
mechanically is questionable, the three components of $m\bf{v}=\bf{p}-\bf{A}$ do not commute,
how can they be measured simultaneously? Second we have to distinguish what is in principle
not measurable and what is not measurable at present.
M. Wakamatsu argued that the quark canonical OAM in nucleon might be not measurable.
Yes, up to now we don't know how to measure it. But we suppose the same canonical OAM
of electron in atom is measurable. Usually we believe the multipole radiation analysis
gives us the information about the electron canonical OAM in atom, even the quark canonical OAM
in nucleon. In this sense we don't think the relevant decompositions of nucleon spin listed
in L.L.'s review paper is complete. The momentum, spin and OAM proposed by us is in principle
measurable and we'd better do not close the door to explore the measuring method of these observables.

Certainly there are more problems which we have not discussed here. We also apologize for
that we have not discussed many interesting developments contributed by other authors
in this field due to space limitation.

This work is supported by NSFC grant 11035003,11035006,11175088,11175215,and 11275077.
It is an invited contribution to the Special Issue on Path Dependence in Quantum Physics, 
to be published in Physics of Elementary Particles and Atomic Nuclei.

\end{document}